\documentclass{moriond}
\usepackage{amssymb}
\usepackage{amsmath}
\usepackage{color}

\def\be{\begin{equation}}
\def\ee{\end{equation}}
\def\bea{\begin{eqnarray}}
\def\eea{\end{eqnarray}}

\def\ahvp{a_\mu^{\rm hvp}}
\def\ahvplo{a_\mu^{\rm hvp,\,LO}}
\def\ahlbl{a_\mu^{\rm hlbl}}
\def\awin{a_\mu^{\rm win}}

\newcommand{\Photo}{}

\begin{document}

\vspace*{4cm}
\title{Progress on $(g-2)_\mu$ from Lattice QCD}

\author{Hartmut Wittig}

\address{Institute for Nuclear Physics, PRISMA$^+$ Cluster of Excellence and Helmholtz Institute Mainz\\
Becher Weg 45, D-55099 Mainz, Germany}
\maketitle\abstracts{ I review the status of lattice QCD calculations
  of the hadronic contributions to the muon's anomalous magnetic
  moment, focussing on the hadronic vacuum polarisation contribution
  which dominates the uncertainty of the Standard Model
  prediction. This quantity exhibits a tension between recent lattice
  QCD results and the traditional data-driven dispersive method. I
  discuss the implications for the running of the electromagnetic
  coupling and the consistency of global fits using electroweak
  precision data.}

\section{Introduction}

The anomalous magnetic moment, $a_\ell$, of a lepton $\ell$
parameterises the fraction of the lepton's interaction strength with a
magnetic field due to quantum corrections. Lepton anomalous magnetic
moments are sensitive probes of the Standard Model and play a pivotal
role in the quest for new physics that may be able to explain the dark
matter puzzle or the observed disparity between matter and antimatter.
The observation of a non-zero deficit between experiment and SM
prediction would be attributed to physics beyond the SM, i.e.
\begin{equation}
    a_\ell^{\rm exp}-(a_\ell^{\rm QED}+a_\ell^{\rm weak}+a_\ell^{\rm strong}) = a_\ell^{\rm BSM}\,.
\vspace{-0.1cm}
\end{equation}
The muon anomalous magnetic moment is a particularly promising quantity due to the fact that the BSM contribution scales like
%
    $a_\ell^{\rm BSM}\propto m_\ell^2/M_{\rm BSM}^2$,
%
where $m_\ell$ is the lepton mass and $M_{\rm BSM}$ the BSM
scale. Hence, the sensitivity of $a_\mu$ is enhanced relative to $a_e$
by a factor $(m_\mu/m_e)^2\approx 4\cdot10^4$. Moreover, $a_\mu$ can
be measured with a precision of 0.46\,ppm \cite{Muong-2:2021ojo} which
is much more precise than what can currently be achieved
experimentally for $a_\tau$, which would be even more sensitive.

A consensus value for the SM prediction $a_\mu^{\rm SM}\equiv
a_\mu^{\rm QED}+a_\mu^{\rm weak}+a_\mu^{\rm strong}$ has been reported
in a 2020 White Paper by the Muon $g-2$ Theory
Initiative.\cite{Aoyama:2020ynm} The quoted overall precision of 0.37
ppm is limited by the strong interaction, notably the contributions
from hadronic vacuum polarisation (HVP) and light-by-light scattering
(HLbL), i.e. $a_\mu^{\rm strong}=\ahvp+\ahlbl$. By far the biggest
share of the theory error (i.e. 83\%) is due to $\ahvp$ while $\ahlbl$
accounts for almost 17\% of the uncertainty. The White
Paper-recommended value for $a_\mu^{\rm SM}$ is based on the
``data-driven'' evaluation of the HVP contribution in terms of
dispersion integrals and experimentally measured hadronic cross
sections. It exhibits a sizeable tension of $4.2$ standard deviations
with the current experimental average
\begin{equation}\label{eq:WP}
    a_\mu^{\rm exp}-a_\mu^{\rm SM}=(25.1\pm5.9)\cdot10^{-10}\,,
\vspace{-0.1cm}
\end{equation}
which is tantalisingly close to the $5\sigma$ threshold required to
claim a quantitative failure of the SM. Lattice QCD provides a viable
alternative to the data-driven method.\cite{DellaMorte:2017dyu,Chakraborty:2017tqp,Borsanyi:2017zdw,Blum:2018mom,Giusti:2019xct,Shintani:2019wai,FermilabLattice:2019ugu,Gerardin:2019rua,Aubin:2019usy,Giusti:2019hkz,Borsanyi:2020mff,Lehner:2020crt,Aubin:2022hgm}
As of today, however, only the BMW collaboration
\cite{Borsanyi:2020mff} has published a lattice result for the
leading-order HVP contribution $\ahvplo$ with similar precision
compared to the data-driven method. If one were to replace $\ahvplo$
in the White Paper (WP) by BMW's estimate, the difference to the
experimental average would be reduced to just $1.5\sigma$, i.e.
\begin{equation}\label{eq:WP2BMW}
    a_\mu^{\rm exp}-\left.a_\mu^{\rm SM}\right|_{\rm WP\to BMW}=(10.7\pm7.0)\cdot10^{-10}\,.
\vspace{-0.1cm}
\end{equation}
At the same time, the result is in tension with the data-driven value
at the level of $2.1\sigma$. Clearly, this requires an independent
validation.
By contrast, the situation regarding the hadronic light-by-light
scattering contribution is quite stable, with lattice
calculations \cite{Blum:2019ugy,Chao:2021tvp,Chao:2022xzg,Blum:2023vlm}
in good agreement with phenomenological
estimates.\cite{melnikov:2003xd,masjuan:2017tvw,Colangelo:2017fiz,hoferichter:2018kwz,gerardin:2019vio,bijnens:2019ghy,colangelo:2019uex,pauk:2014rta,danilkin:2016hnh,jegerlehner:2017gek,knecht:2018sci,eichmann:2019bqf,roig:2019reh}

\vspace{-0.1cm}
\section{Hadronic vacuum polarisation contribution}
\vspace{-0.1cm}
\subsection{The data-driven approach}
\vspace{-0.1cm} The standard method to determine the leading-order HVP
contribution $\ahvplo$ is based on the evaluation of a dispersion
integral over the so-called $R$-ratio $R(s)$, i.e. the normalised
total hadronic cross section for $e^+e^-\to\rm hadrons$, times a
slowly varying kernel function $\hat{K}(s)\approx1$:
\begin{equation}\label{eq:Rratio}
    \ahvplo=\left(\frac{\alpha m_\mu}{3\pi}\right)^2\int_{m_{\pi^0}^2}^\infty
    \frac{\hat{K}(s)}{s^2} R(s)\,{\rm d}s\ ,\quad
    R(s)=\frac{3s}{4\pi\,\alpha^2}\,\sigma(e^+e^-\to{\rm hadrons})\,.
\vspace{-0.1cm}
\end{equation}
Owing to the factor of $s^2$ in the denominator, the integral receives
its main contribution from the low-energy region, in particular from
the dominant channel $e^+e^-\to\pi^+\pi^-$. Since perturbation theory
in the strong coupling $\alpha_s$ cannot be applied near the pion
threshold, one has to resort to experimental measurements of the
$R$-ratio to evaluate the integral in eq.\,(\ref{eq:Rratio}), implying
that the resulting theoretical prediction is subject to experimental
uncertainties. Indeed, there is a long-standing tension in the
dominant two-pion channel between the measurements of the KLOE
\cite{KLOE-2:2017fda} and BaBar \cite{BaBar:2012bdw} collaborations.
The WP-recommended value takes this into account by quoting a separate
systematic error in addition to the experimental error
\begin{equation}\label{eq:HVPdata}
    \ahvplo=693.1(2.8)_{\rm exp}\,(2.8)_{\rm syst}\,(0.7)_{\rm DV+QCD}\cdot10^{-10} = (693.1\pm4.0)\cdot10^{-10}\,,
\vspace{-0.1cm}
\end{equation}
which also accounts for differences in the analyses of cross-section
data\,\cite{Davier:2017zfy,Keshavarzi:2018mgv,Colangelo:2018mtw,Hoferichter:2019mqg,Davier:2019can,Keshavarzi:2019abf}
while the third quoted error is related to theoretical
uncertainties. Adding the individual errors in quadrature results in a
total precision of 0.6\%. However, the recent measurement of the
dominant two-pion channel by the CMD-3
experiment\,\cite{CMD-3:2023alj} has made the situation considerably
more complicated: not only does it disagree with the earlier result
from the predecessor experiment CMD-2 but also with all other
measurements published prior to 2023. Taken at face value, the CMD-3
result is consistent with the higher estimate for $\ahvplo$ suggested
by lattice QCD. However, the tension among the $e^+e^-$ data has
increased significantly and must be understood.

\vspace{-0.1cm}
\subsection{The HVP contribution in lattice QCD}
\begin{figure}[t]
    \centering
    \includegraphics[width=0.98\textwidth]{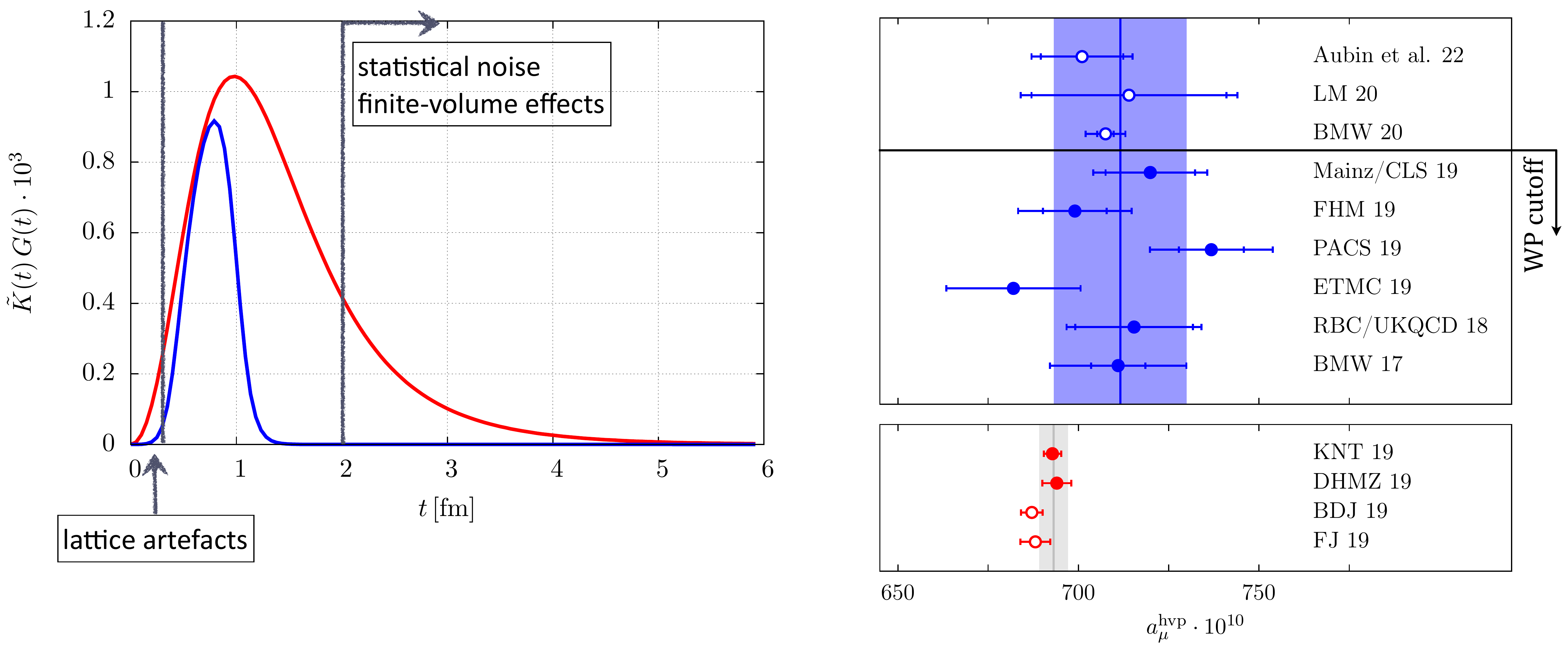}
    \vspace{-0.25cm}
    \caption{{\bf Left:} The TMR integrand, defined in
      Eq.\,(\ref{eq:TMRdef}) plotted versus Euclidean time (red
      curve). The blue curve represents the corresponding integrand of
      the intermediate window observable $\awin$. {\bf Right:}
      Compilation of results for $\ahvplo$. Blue circles represent
      lattice results
      \protect\cite{Borsanyi:2017zdw,Blum:2018mom,Giusti:2019xct,Shintani:2019wai,FermilabLattice:2019ugu,Gerardin:2019rua,Borsanyi:2020mff,Lehner:2020crt,Aubin:2022hgm}
      while red circles denote results obtained from the
      $R$-ratio.\protect\cite{Keshavarzi:2019abf,Davies:2019efs,Benayoun:2019zwh,Jegerlehner:2017lbd}
      The blue band represents the average over lattice results
      represented by the full blue circles, as quoted in the 2020
      WP. The grey band is the estimate of
      Eq.\,(\ref{eq:HVPdata}). Results denoted by open circles were
      published after the release of the WP.}
    \label{fig:HVPlat}
\end{figure}

\vspace{-0.1cm} 
The lattice approach to determining $\ahvplo$ differs substantially
from the data-driven method. For once, lattice QCD does not compute
the $R$-ratio from first principles. Rather, the value of $\ahvplo$
is obtained from the spatially summed correlator $G(t)$ of the
electromagnetic current $J_\mu^{\rm em}(x)$ convoluted with an
analytically known function $\tilde{K}(t)$ and integrated over the
Euclidean time variable $t$, i.e.\,\cite{Bernecker:2011gh}
\begin{equation}\label{eq:TMRdef}
    \ahvplo=\Big(\frac{\alpha}{\pi}\Big)^2\int_0^{\infty}dt\,\tilde{K}(t)\,G(t),\quad G(t)\delta_{kl}=-a^3\sum_{\vec{x}}\left\langle J^{\rm em}_k(x)\,J^{\rm em}_l(0) \right\rangle\,.
\vspace{-0.1cm}
\end{equation}
This expression, known as the ``time-momentum representation" (TMR)
allows for an inclusive determination of $\ahvplo$ and is not
sensitive to individual hadronic channels. The integrand
$\tilde{K}(t)G(t)$ is shown in the left panel of
Fig.\,\ref{fig:HVPlat}. The evaluation of the TMR integral in lattice
QCD does not rely on experimental data, except for simple input
quantities such as hadron masses to set the scale and calibrate the
quark masses. However, the goal of determining $\ahvplo$ with
sub-percent precision presents several challenges for lattice
calculations. The first is related to the tail of the integrand,
i.e. the region of $t\gtrsim2\,\rm fm$, which contributes about 3\% to
the value of $\ahvplo$ but is subject to strong statistical
fluctuations due to the exponentially increasing statistical noise in
the correlator $G(t)$ as $t\to\infty$. The large-$t$ regime also
contributes the bulk of finite-volume effects which the data must be
corrected for. The region of small Euclidean distances is most sensitive
to discretisation effects (``lattice artefacts") which must be removed
via a careful extrapolation to the continuum limit. This is
increasingly challenging for sub-percent statistical precision, since
it requires the ability to disentangle a complicated pattern of terms
beyond the leading corrections in the lattice
spacing.\cite{Husung:2019ytz,Ce:2021xgd,Husung:2022kvi} Finally, for
sub-percent precision, isospin-breaking effects arising from unequal
up- and down-quark masses and electromagnetism must be accounted for.

The chosen discretisation of the quark action has a major influence on
the quality and cost of lattice calculations for $\ahvplo$. Among the
most widely used quark actions are rooted staggered fermions,
Wilson-type quarks,
and domain wall fermions.
A detailed overview of different actions is presented in Appendix~A.2
of the review.\cite{Meyer:2018til} Here we only mention that rooted
staggered quarks are subject to sizeable lattice artefacts from
remnant spurious degrees of freedom called ``tastes", which must be
removed analytically before results are extrapolated to vanishing
lattice spacing. Wilson-type quarks do not require such corrections
but are not rigorously protected against the appearance of small or
negative eigenvalues of the Wilson-Dirac operator, as a result of
explicit chiral symmetry breaking. This not only increases the
numerical cost but also makes it harder to simulate at the physical
value of the pion mass. The domain wall action describes a single
quark flavour and preserves chiral symmetry up to exponentially small
corrections. This comes at the significant expense of having to
simulate the theory on a five-dimensional lattice.

The current set of avaliable lattice results for $\ahvplo$ is shown in
the right upper panel of Fig.\,\ref{fig:HVPlat}. The blue band
represents the average of peer-reviewed, published results at the time
of the release of the WP. While the central value is higher than the
estimate from the data-driven method of Eq.\,(\ref{eq:HVPdata}), the
relative uncertainty of 2.6\% is larger by more than a factor four.
Given that the discretisations of the quark action and and the
procedures employed to produce the lattice results shown in
Fig.\,\ref{fig:HVPlat} are quite different, it is remarkable and
highly non-trivial that they agree at this level of precision.
The result labelled RBC/UKQCD\,18\,\cite{Blum:2018mom} has been
obtained on two ensembles of domain wall quarks, with lattice spacings
of $a=0.114$\,fm and $0.084$\,fm, both directly at the physical pion
mass. The leading isospin-breaking corrections of about 1\% have been
computed and added to the result before extrapolating the results to
the continuum limit assuming an {\it ansatz} proportional to $a^2$,
augmented by an O($a^4$)-term. The precision at the physical point is
2.6\%. The calculation by Mainz/CLS\,\cite{Gerardin:2019rua} is based
on a total of 15 ensembles of O($a$)-improved Wilson fermions,
covering four values of the lattice spacing between
$a=0.085-0.050$\,fm and pion masses in the range $m_\pi=130-420$\,MeV.
In order to account for the neglected isospin-breaking effects, the
size of the isospin-breaking correction computed by the ETM
collaboration\,\cite{Giusti:2019xct} has been added to the error of
the final result which has a total uncertainty of 2.2\%. The BMW
collaboration\,\cite{Borsanyi:2020mff} computed $\ahvplo$ using 27
ensembles of rooted staggered quarks at six values of the lattice
spacing between $a=0.132-0.064$\,fm, all at the physical pion mass.
Several different models were applied to remove taste-induced lattice
artefacts prior to performing extrapolations in the lattice spacing to
the continuum limit. The final result is selected from a distribution
of different fits. After correcting for strong and electromagnetic
isospin-breaking effects, BMW quote a value of
$\ahvplo=(707.5\pm2.3\pm5.0)\cdot10^{-10}$, which has a total
precision of 0.8\%, only slightly worse than that of the data-driven
result. This result, when supplied for the HVP contribution rather
than the estimate of Eq.\,(\ref{eq:HVPdata}), is responsible for the
reduction of the tension between SM prediction and experimental
average in Eq.\,(\ref{eq:WP2BMW}).

\vspace{-0.3cm}
\subsection{Window observable}
\vspace{-0.1cm}

\begin{figure}[t]
    \centering
    \includegraphics[width=0.6\textwidth]{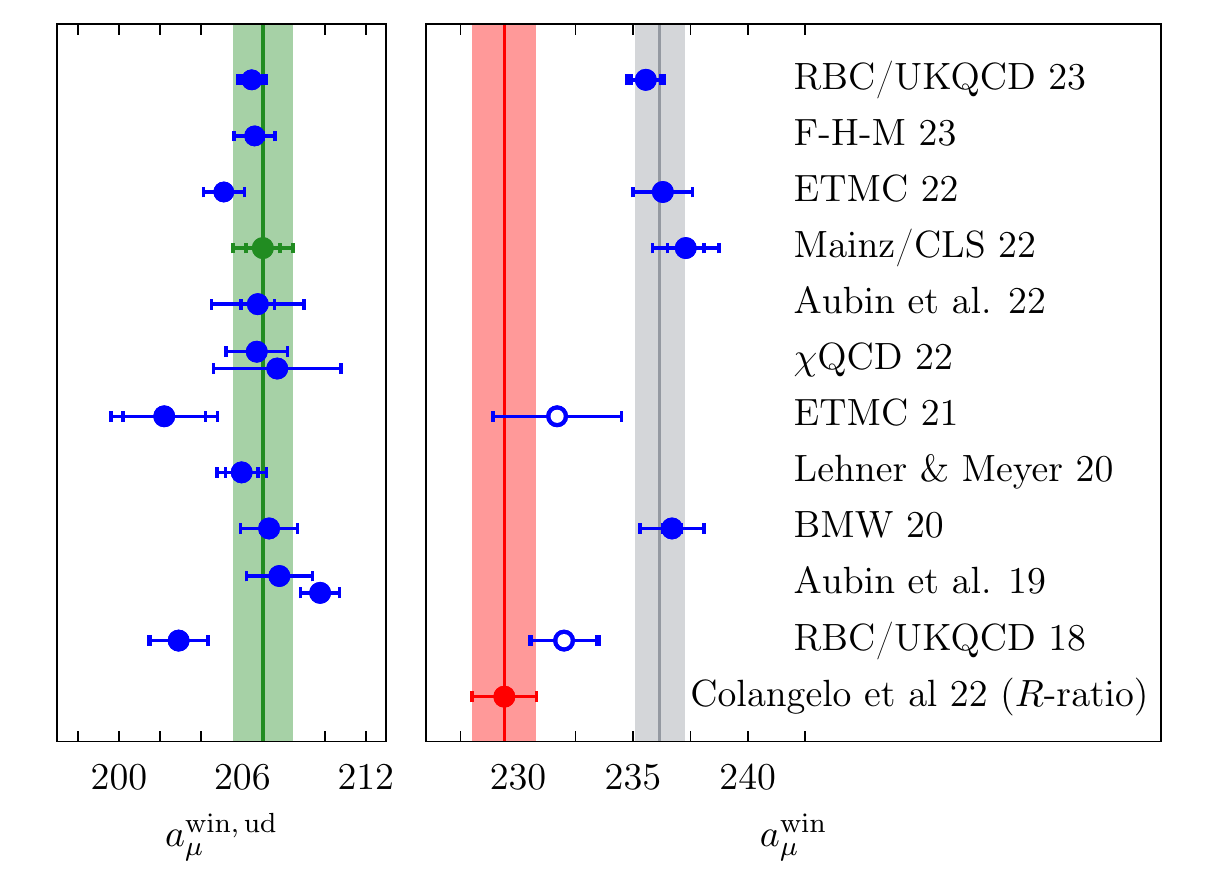}
    \vspace{-0.3cm}
    \caption{Lattice results for the window observable
      \protect\cite{Blum:2023qou,Bazavov:2023has,ExtendedTwistedMass:2022jpw,Ce:2022kxy,Aubin:2022hgm,Wang:2022lkq,Giusti:2021dvd,Lehner:2020crt,Borsanyi:2020mff,Aubin:2019usy,Blum:2018mom}
      in units of $10^{-10}$. The left panel shows results for the
      dominant contribution from $u,\,d$ quarks. The green vertical
      band denotes the result from Mainz/CLS.\protect\cite{Ce:2022kxy}
      Adding the contributions from strange and charm quarks, as well
      as disconnected contributions and isospin-breaking corrections
      yields the results shown in the right panel. The red data point
      and red vertical band represents the recent data-driven estimate
      \protect\cite{Colangelo:2022vok} for the window observable. The
      global average over the solid blue points, represented by the
      grey band, is in tension with the data-driven estimate by
      $3.8\sigma$. }
    \label{fig:window}
\end{figure}

A cross-check of the BMW result with sub-percent precision can be
performed by restricting the integration in the TMR integral,
Eq.\,(\ref{eq:TMRdef}), to a subinterval which essentially removes the
regions of strong statistical fluctuations, large finite-volume
effects and large lattice artefacts. This is the idea behind the
so-called ``window observables", first introduced by
RBC/UKQCD.\cite{Blum:2018mom} More specifically, the ``intermediate
window observable" $\awin$ is obtained via a convolution of the TMR
integrand with an additional weight function $W(t;\,t_0,\,t_1)$
according to
\begin{equation}\label{eq:window}
    \awin=\Big(\frac{\alpha}{\pi}\Big)^2\int_0^{\infty}dt\,\tilde{K}(t)\,G(t)\,W(t;\,t_0,\,t_1)\,,\quad 
    W(t;\,t_0,\,t_1)=\Theta(t,\,t_0,\,\Delta)-\Theta(t,\,t_1,\,\Delta)\,,
\vspace{-0.1cm}
\end{equation}
where the smoothed step function $\Theta(t,\,t',\,\Delta)$ is defined
as $\Theta(t,\,t',\,\Delta)=\frac{1}{2}[1+\tanh(t-t')/\Delta]$.
Choosing $t_0=0.4$\,fm, $t_1=1.0$\,fm and the width as
$\Delta=0.15$\,fm yields the blue curve in the left panel of
Fig.\,\ref{fig:HVPlat}. With this choice one finds that finite-volume
corrections to $\awin$ are reduced to just 0.25\% and that the total
uncertainty is dominated by statistics. This makes the intermediate
window observable an ideal benchmark quantity for comparing different
lattice calculations. In addition, it is possible to evaluate $\awin$
using the $R$-ratio. Indeed, motivated by the procedure used to
determine the WP estimate for $\ahvplo$ from $e^+e^-$ cross section
data published prior to 2023, one finds\,\cite{Colangelo:2022vok}
\begin{equation}\label{eq:WINdata}
    \left.\awin\right|_{R{\rm-ratio}}=(229.4\pm1.4)\cdot10^{-10}\,.
\vspace{-0.1cm}
\end{equation}
The collection of lattice results for the window observable is shown
in Fig.\,\ref{fig:window}.
From the compilation shown in the left panel one concludes that
lattice QCD produces consistent results for the dominant light-quark
connected contribution for a wide range of different discretisations
and with sub-percent precision. The only exceptions are the
calculations labelled RBC/UKQCD\,18~\cite{Blum:2018mom} and
ETMC\,21~\cite{Giusti:2021dvd} which have since been superseded by
RBC/UKQCD~23\,\cite{Blum:2023qou} and
ETMC\,22.\cite{ExtendedTwistedMass:2022jpw} After adding the
contributions to $\awin$ from strange and charm quarks, as well as
quark-disconnected contributions and isospin-breaking corrections, one
obtains the results shown in the right panel, which can be readily
compared to the data-driven estimate of Eq.\,(\ref{eq:WINdata}). One
observes a clear tension between the most recent lattice estimates and
the corresponding result extracted from $e^+ e^-$ hadronic cross
sections. In order to arrive at a more quantitative statement, I have
performed a global average of the results labelled BMW\,20
\cite{Borsanyi:2020mff}, Mainz/CLS\,22 \cite{Ce:2022kxy}, ETMC\,22
\cite{ExtendedTwistedMass:2022jpw} and RBC/UKQCD\,23
\cite{Blum:2023qou} assuming
that the results are 100\% correlated. In this way one obtains a
tentative lattice average of $\awin=(236.2\pm1.1)\cdot10^{-10}$, which
differs from Eq.\,(\ref{eq:WINdata}) by $3.8\sigma$, i.e.
\begin{equation}
    \left.\awin\right|_{\rm Lattice}-\left.\awin\right|_{R{\rm-ratio}}=(6.8\pm1.8)\cdot10^{-10}\,.
\vspace{-0.1cm}
\end{equation}
Thus, there is a confirmed tension between lattice QCD and $e^+ e^-$
data prior to 2023 (i.e. excluding the recent CMD-3 result). It is
also interesting to note that the intermediate window accounts for
50\% of the tension between the $R$-ratio estimate and the BMW result
for $\ahvplo$, as can be easily seen by comparing Eqs.\,(\ref{eq:WP})
and\,(\ref{eq:WP2BMW}). One can go one step further and study the
implications of the confirmed tension in the window observable for the
SM prediction of $a_\mu$ and the direct measurement. Since the
$R$-ratio estimate for $\awin$ is based on the same procedure as the
WP-recommended value for $\ahvplo$, I have performed the exercise of
subtracting Eq.\,(\ref{eq:WINdata}) from the SM prediction and
replacing it by the global average of lattice results for
$\awin$. This reduces the discrepancy between the SM prediction and
the experimental average to just over three standard deviations, i.e.
\begin{equation}
    a_\mu^{\rm exp}-\left.a_\mu^{\rm SM}\right|_{R{\rm-ratio}\to {\rm Lattice}}^{\rm win}=(18.3\pm5.9)\cdot10^{-10}\qquad [3.1\sigma]\,.
\vspace{-0.1cm}
\end{equation}
%
A more thorough and complete analysis will be presented in an update
of the 2020 WP, which is currently being prepared by the Muon $g-2$
Theory Initiative.

Tracing the origin of the tension between lattice QCD and the
data-driven approach is obviously a burning issue. One proposal
\cite{FermilabLattice:2022izv} stresses the role of window observables
defined for alternative parameter choices for $t_0,\,t_1$ and
$\Delta$. Alternatively, attempts have been made to determine the
spectral function $R(s)_{\rm lat}$ associated with the vector
correlator $G(t)$ and compare it to the $R$-ratio $R(s)_{e^+ e^-}$
measured in $e^+ e^-$ annihilation.\cite{Alexandrou:2022tyn} It is
clearly a crucial question in which energy range the two spectral
functions differ. Owing to the restriction of the integration to the
interval between 0.4\,fm and 1.0\,fm, the two-pion channel, which
contributes about 70\% to the value of $\ahvplo$, is less dominant in
the window observable $\awin$. An interesting observation in this
context was made by Mainz/CLS\,\cite{Ce:2022kxy} based on a
phenomenological model: Namely, if the spectral function $R(s)_{\rm
  lat}$ associated with $G(t)$ were somehow enhanced by an amount
$\epsilon$ in the interval $\sqrt{s}=600-900$\,MeV relative to
$R(s)_{e^+ e^-}$, this would produce an enhancement of $0.6\epsilon$
in both $\ahvplo$ and $\awin$. Taking the ratio of the lattice average
with Eq.\,(\ref{eq:WINdata}) one finds $\awin|_{\rm
  Lattice}/\awin|_{R{\rm-ratio}}=1.030(8)$ which, according to the
above reasoning, implies that $R(s)_{\rm lat}$ would be larger by
about 5\% compared to $R(s)_{e^+ e^-}$ in the region
$\sqrt{s}=600-900$\,MeV.
An enhancement of this magnitude would suggest that the value of
$\ahvplo$ might be even larger than the estimate published by BMW.

\vspace{-0.2cm}
\section{Relation to the hadronic running of the electromagnetic coupling}
\vspace{-0.2cm}
An important consistency check can be performed for the closely
related quantity $\Delta\alpha_{\rm had}(q^2)$ which denotes the
hadronic contribution to the running of the electromagnetic coupling
$\alpha$. Like the HVP contribution to the muon $g-2$,
$\Delta\alpha_{\rm had}$ is accessible via the vector correlator
$G(t)$. For instance, let $-Q^2 < 0$ be a given spacelike momentum
transfer, then $\Delta\alpha_{\rm had}(-Q^2)$ can be computed from the
convolution integral \cite{Bernecker:2011gh}
\begin{equation}
    \Delta\alpha_{\rm had}(-Q^2)=\frac{\alpha}{\pi}\frac{1}{Q^2}\int_0^{\infty} dt\,G(t)\,\Big[Q^2t^2-4\sin^2({\textstyle\frac{1}{2}}Q^2t^2)\Big]\,.
\vspace{-0.1cm}
\end{equation}
Alternatively, one can express $\Delta\alpha_{\rm had}(q^2)$ for some
(spacelike or timelike) momentum transfer $q^2$ via the principal
value of a dispersion integral involving the $R$-ratio, according to
\begin{equation}
    \Delta\alpha_{\rm had}(q^2)=-\frac{\alpha\,q^2}{3\pi}\; {\cal P}\hspace{-12pt}\int_{m_{\pi^0}^2}^{\infty}ds\,\frac{R(s)}{s(s-q^2)}\,.
\vspace{-0.1cm}
\end{equation}
Mainz/CLS have compared the results from a direct lattice calculation
\cite{Ce:2022eix} of $\Delta\alpha_{\rm had}(-Q^2)$ to those from
data-driven evaluations
\cite{Keshavarzi:2019abf,Davier:2019can,Jegerlehner:2019lxt} for
several fixed values of $Q^2$ in the range between $1.0$ and $5.0\,\rm
GeV^2$. Indeed, for $Q^2\gtrsim3\,\rm GeV^2$, lattice and dispersive
estimates differ at the level of $3\sigma$, in complete correspondence
to the tension observed for the window observable $\awin$.
Interestingly, in a recent paper \cite{Davier:2023hhn} it was found
that the Adler function $D(Q^2)$ (which is the derivative of
$\Delta\alpha_{\rm had}(-Q^2)$), computed in massive QCD perturbation
theory at four loops, shows good agreement with $D(Q^2)$ derived from
the lattice calculation of $\Delta\alpha_{\rm had}$, while there is a
tension with the Adler function determined from $e^+ e^-$ data.

\begin{figure}
    \centering
    \includegraphics[width=0.5\textwidth]{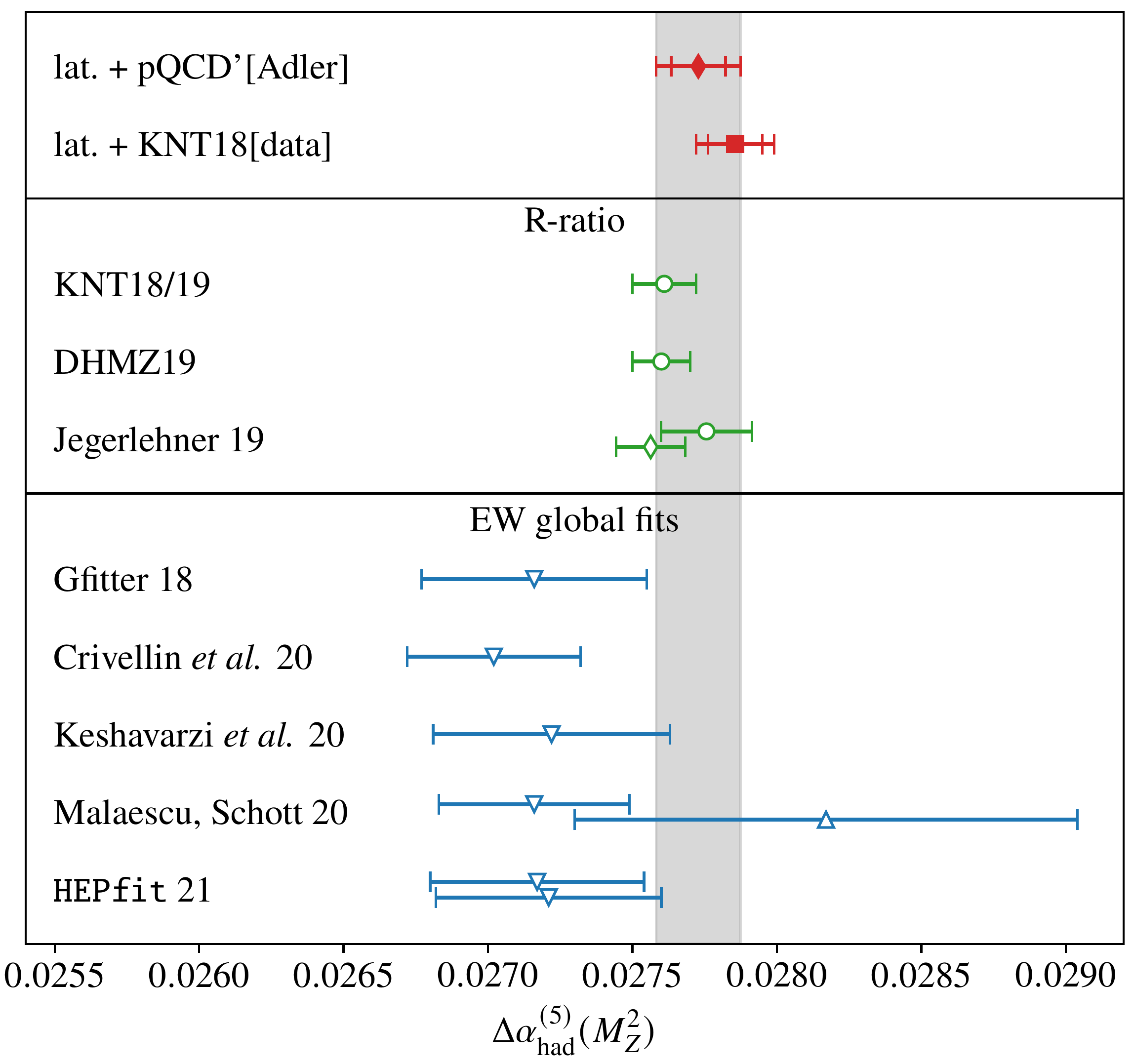}
    \vspace{-0.25cm}
    \caption{Compilation of results for the hadronic running of
      $\alpha$ at the $Z$~pole. Red data points represent results
      obtained via the Euclidean split technique and lattice QCD input
      \protect\cite{Ce:2022eix}, results based on dispersion theory are
      shown as green symbols
      \protect\cite{Keshavarzi:2019abf,Davier:2019can,Jegerlehner:2019lxt}
      while blue symbols represent estimates obtained from global
      electroweak
      fits.\protect\cite{Crivellin:2020zul,Keshavarzi:2020bfy,Malaescu:2020zuc,Haller:2018nnx,deBlas:2021wap}
      The lattice result of Eq.\,(\ref{eq:running-lattice}) is shown
      as the grey vertical band.}
    \label{fig:running}
\end{figure}

Since $\ahvplo$ and $\Delta\alpha_{\rm had}$ are directly correlated,
one might think that an enhancement of both quantities as suggested by
lattice QCD would produce a tension with the prediction of the
hadronic running of $\alpha$ from global fits to electroweak precision
data. To see whether this is the case, we must convert the lattice
estimate for $\Delta\alpha_{\rm had}(-Q^2)$ into an estimate at the
$Z$~pole. This can be done reliably using the ``Euclidean split
technique'', in which $\Delta\alpha_{\rm had}^{(5)}(M_Z^2)$ is divided
into three separate contributions, according to \cite{Jegerlehner:2008rs}
\bea
    \Delta\alpha_{\rm had}(M_Z^2) &=&  \Delta\alpha_{\rm had}(-Q_0^2) \nonumber\\
    &+& [\Delta\alpha_{\rm had}(-M_Z^2)-\Delta\alpha_{\rm had}(-Q_0^2)] +
        [\Delta\alpha_{\rm had}(M_Z^2) -\Delta\alpha_{\rm had}(-M_Z^2)]\,.\label{eq:EuclSplit}
\eea
To determine $\Delta\alpha_{\rm had}(M_Z^2)$ one can use the lattice
result for $\Delta\alpha_{\rm had}(-Q_0^2)$ by Mainz/CLS at
$Q_0^2=5\,\rm GeV^2$ as input, while the two terms in square brackets
on the right-hand side can be evaluated in QCD perturbation
theory. This yields the result~\cite{Ce:2022eix}
\begin{equation}\label{eq:running-lattice}
    \Delta\alpha_{\rm had}(M_Z^2)=0.027\,73(9)_{\rm lat}(2)_{\rm btm}(12)_{\rm pQCD}\,,
\vspace{-0.1cm}
\end{equation}
where the first error is the total uncertainty of the lattice
calculation, the second error quantifies the neglected contribution
from bottom quarks, and the third error accounts for the uncertainty
of the perturbative running and matching. In Fig.\,\ref{fig:running}
the above result, shown as the grey vertical band, is compared to
dispersion theory and global electroweak fits. Clearly, the results
are compatible within errors, which signals that larger values of
$\ahvplo$ and $\Delta\alpha_{\rm had}(M_Z^2)$ are not excluded by
electroweak precision data. At first sight, the agreement of
$\Delta\alpha_{\rm had}(M_Z^2)$ between lattice QCD and dispersion
theory in Fig.\,\ref{fig:running} seems to contradict the earlier
statement that a $3\sigma$-discrepancy is observed at low Euclidean
momentum transfers. However, the resolution of this apparent paradox
is obtained by realising that the same running and matching factors
are applied when converting $\Delta\alpha_{\rm had}$ at low $Q^2$ to
the corresponding estimate at the $Z$~pole. The associated
uncertainties in the running cancel in the correlated difference
between lattice QCD and dispersion theory, thereby preserving the
tension.\cite{SanJosePerez:2022qwc}

\vspace{-0.2cm}
\section{Summary and outlook}
\vspace{-0.25cm} An unambiguous interpretation of the new measurement
of the muon $g-2$ by the E989 experiment at Fermilab
\cite{Muong-2:2021ojo} is impeded by several tensions that have been
exposed since the publication of the 2020 White Paper: (1) There is a
tension of $2.1\sigma$ between a single lattice calculation
\cite{Borsanyi:2020mff} and the WP-recommended value for $\ahvplo$,
based on $e^+ e^-$ cross section data published prior to 2023; (2)
There is a tension of almost $4\sigma$ between several lattice
calculations
\cite{Borsanyi:2020mff,Ce:2022kxy,ExtendedTwistedMass:2022jpw,Blum:2023qou}
and the corresponding dispersive estimate \cite{Colangelo:2022vok}
based on the same $e^+ e^-$ data; (3) There is a tension of
$2-3\sigma$ in the hadronic running of $\alpha$, as estimated by two
lattice calculations \cite{Borsanyi:2020mff,Ce:2022eix} and $e^+ e^-$
data; (4) There is a slight tension of $1-2\sigma$ in the Adler
function determined from lattice and perturbative QCD on the one hand,
and $e^+ e^-$ data on the other; (5) Finally, there is a tension of
$2.7\sigma$ in the dominant $\pi^+\pi^-$ channel between
BaBar~\cite{BaBar:2012bdw} and KLOE~\cite{KLOE-2:2017fda}, as well as
a tension of about $4\sigma$ between CMD-3 \cite{CMD-3:2023alj} and
all other experiments. In this context, it is important to realise
that a larger SM prediction for $a_\mu$ is not in contradiction with
global electroweak constraints, at least at the current level of
precision.
Obviously, an independent cross-check of the BMW lattice result for
$\ahvplo$ with sub-percent precision is badly needed. Furthermore, the
tension among $e^+ e^-$ data must be elucidated, a task for which the
alternative determination of the $R$-ratio from $\tau$ decays could be
useful.\cite{Masjuan:2023qsp}
These activities are currently in progress.
The Muon $g-2$ Theory Initiative is preparing an update of the
original WP, which will thoroughly address the issues that have come
to the fore since 2020.

\medskip\par\noindent{\bf Acknowledgments:}
It is a pleasure to thank the organisers for the invitation to La
Thuile. I am grateful to Achim Denig, Harvey Meyer and Toni Pich for
stimulating discussions, and to Aida El Khadra and Martin Hoferichter
for useful comments. This work is partially supported by the Cluster
of Excellence Precision Physics, Fundamental Interactions, and
Structure of Matter (PRISMA+ EXC 2118/1) funded by DFG within the
German Excellence Strategy (Project ID 39083149). Calculations by
Mainz/CLS were performed on HPC platforms at Mainz, JSC J\"ulich and
HLRS Stuttgart. The support of the Gauss Centre for Supercomputing
(GCS) and the John von Neumann-Institut für Computing (NIC) for
projects CHMZ21 and CHMZ23 at JSC and project GCS-HQCD at HLRS is
gratefully acknowledged.

\vspace{-0.2cm}
\section*{References}
\vspace{-0.2cm}


\providecommand{\href}[2]{#2}\begingroup\raggedright\endgroup

\end{document}